\documentclass[twocolumn]{revtex4}
\usepackage{graphicx}
\usepackage{amsmath,amsbsy,amssymb}
\usepackage{bm}
\usepackage{mathrsfs}
\usepackage{ulem}

\newcommand{\diff}{\mathrm{d}}
\newcommand{\imag}{\mathrm{Im}\,}

\newcommand{\imu}{\mathrm{i}}
\newcommand{\epn}{\mathrm{e}}

\newcommand{\ua}{\uparrow}
\newcommand{\da}{\downarrow}
\newcommand{\dg}{\dagger}
\newcommand{\la}{\langle}
\newcommand{\ra}{\rangle}

\newcommand{\sg}{\sigma}
\newcommand{\gm}{\gamma}
\newcommand{\ep}{\varepsilon}

\begin{document}

\title{
Spontaneous orbital-selective Mott transitions and the Jahn-Teller metal of A$_3$C$_{60}$
}

\author{Shintaro Hoshino$^{1}$ and Philipp Werner$^2$}

\affiliation{
$^1$RIKEN Center for Emergent Matter Science (CEMS), Wako, 351-0198 Saitama, Japan
\\
$^2$Department of Physics, University of Fribourg, 1700 Fribourg, Switzerland
}

\date{\today}

\begin{abstract}
The alkali-doped fullerides A$_3$C$_{60}$ are half-filled three-orbital Hubbard systems which exhibit an unconventional superconducting phase next to a Mott insulator. While the pairing is understood to arise from an effectively negative Hund coupling, the highly unusual Jahn-Teller metal near the Mott transition, featuring both localized and itinerant electrons, has not been understood. This property is consistently explained by a previously unrecognized phenomenon: the spontaneous transition of multiorbital systems with negative Hund coupling into an orbital-selective Mott state. This symmetry-broken state, which has no ordinary orbital moment, is characterized by an orbital-dependent two-body operator (the double occupancy) or an orbital-dependent kinetic energy, and may be regarded as a diagonal-order version of odd-frequency superconductivity. We propose that the recently discovered Jahn-Teller metal phase of Rb$_x$Cs$_{3-x}$C$_{60}$ is an experimental realization of this novel state of matter. 
\end{abstract}

\maketitle

The appearance of long-range order by spontaneous symmetry breaking (SSB) is a fundamental and widely studied concept in physics. In condensed matter systems, the ordered state is typically characterized by an order parameter measuring the charge, magnetic moment, orbital angular momentum, or the pair amplitude in a superconductor. There may however exist more 
complex types
of ordering phenomena.
Here we demonstrate that in a certain class of multi-orbital systems 
one observes an orbital symmetry breaking into a state without conventional order parameter, but with an orbital-dependent double occupancy (a composite order parameter) and an orbital-dependent kinetic energy. 
The 
resulting 
ordered phase is a spontaneous orbital-selective
Mott (SOSM)  state, in which itinerant and localized electrons coexist. 
As this state combines properties of the 
metal (connected to 
the weak-interaction limit) and Mott insulator (connected to 
the strong-interaction limit)
it cannot be detected by perturbative methods from either limits, and its study requires the use of sophisticated techniques.
The 
ordering phenomenon can be discussed in terms of a symmetry-breaking field or order parameter with odd time (frequency) dependence, and is hence related to the concept of odd-frequency superconductivity \cite{berezinskii1974,kirkpatrick1991,balatsky1992,emery1992}.

We will argue that this unconventional SOSM state is realized in alkali-doped fullerides \cite{hebard1991,rosseinsky1991,holczer1991,tanigaki1991,fleming1991,varma1991,takabayashi2009,zadik2015,mitrano2016}, which are promising candidates for diagonal odd-frequency orders.  
Several compounds in this class of materials
 can be regarded as strongly correlated systems since a Mott transition occurs as a function of pressure.
In the case of an fcc lattice \cite{zadik2015}, the Mott insulator stays paramagnetic in a wide range of temperature due to geometrical frustration, 
while it is antiferromagnetically ordered in the case of a bcc lattice, 
which is bipartite \cite{takabayashi2009}. 
With increasing pressure, the system turns into a paramagnetic metal
and is unstable to superconductivity below a maximum $T_c \simeq 38$ K \cite{takabayashi2009}.
Recently, a so-called Jahn-Teller metal (JTM) state has been experimentally identified in fcc 
Rb$_x$Cs$_{3-x}$C$_{60}$ \cite{zadik2015}. This state, which connects the insulating and superconducting phases, exhibits a coexistence of localized and itinerant electrons, but  
the physical origin of this unconventional
feature  
has not been clarified.

An unusual property of fulleride compounds is the effectively negative Hund coupling $J$ produced by anisotropic (Jahn-Teller) phonons \cite{fabrizio1997,capone2002,nomura2012,nomura2015,nomura2015prb,capone2009,nomura2016,hoshino2016,kim2016}.
In contrast to transition metal ($d$-electron) systems, which have a positive Hund coupling of the order of 1 eV, the bare Hund coupling in fullerides is much smaller ($\sim$ 30 meV) because of the spatially 
extended 
molecular orbitals, so that the screening by phonons can lead to a sign inversion of the effective 
$J$ \cite{fabrizio1997,capone2002,nomura2015}.

Theoretically, these systems can be described by a half-filled 
three-orbital Hubbard model 
$\mathscr{H} = \mathscr{H}_\text{kin}+ \mathscr{H}_\text{int}$, 
where
$\mathscr{H}_{\rm kin}= \sum_{\bm k\gm\gm'\sg} 
(\ep_{\bm k\gm\gm'}-\mu \delta_{\gm\gm'}) 
c^\dg_{\bm k\gm\sg} c_{\bm k\gm'\sg}$
is
the kinetic energy of the electrons.
The interaction term is of the form 
\begin{align}
\mathscr{H}_{\rm int} = \sum_{i\gm_1\gm_2\gm_3\gm_4\sg\sg'} U_{\gm_1\gm_2\gm_3\gm_4} c^\dg_{i\gm_1\sg} c_{i\gm_2\sg} c^\dg_{i\gm_3\sg'} c_{i\gm_4\sg'},
\end{align}
 where 
$U_{\gm\gm\gm\gm}=U
/2$ $(>0)$, $U_{\gm\gm\gm'\gm'}=U'
/2$ $(>0)$, $U_{\gm\gm'\gm'\gm}=U_{\gm\gm'\gm\gm'}=J
/2$ $(<0)$ for $\gm\neq \gm'$ and the other components are zero.
We use the standard parametrization $U'=U-2J$ valid for isotropic interactions,
which results in a SU(2)$\times$SO(3) symmetry in spin-orbital space.
The indices $\gm=1,2,3$ and $\sg=\ua,\da$ represent the three degenerate (t$_{1u}$) molecular orbitals and spin, respectively.
We define the spin and orbital moments by
$\bm S_i =\sum_{\gm\sg\sg'} c^\dg_{i\gm\sg} \bm \sg_{\sg\sg'} c_{i\gm\sg'}$
and
$\tau^\eta_i = \sum_{\gm\gm'\sg} c^\dg_{i\gm\sg} \lambda^\eta_{\gm\gm'} c_{i\gm'\sg}$,
where $\bm \sg$ is the Pauli matrix and $\lambda^{\eta=1,\cdots ,8}$ is the Gell-Mann matrix (see Supplemental Material (SM1) for more details 
\cite{supplementary}).
In 
the case 
of the point-group symmetry I relevant 
 for an isolated fullerene, the $\eta=1,3,4,6,8$ (time-reversal even) components, 
which play a central role in the following analysis, 
 correspond to the symmetry H, and $\eta=2,5,7$ (time-reversal odd) to T$_1$ \cite{saito91}. These are analogs of
quadrupole and dipole moments, made from $p$-electrons with angular momentum $\ell=1$.

To solve the lattice model, we use 
the dynamical mean-field theory (DMFT) \cite{kuramoto1985,georges1996} in combination with the continuous-time quantum Monte Carlo method \cite{werner2007matrix}. 
This is a suitable method
for three-dimensional systems with strong local correlations, such as fulleride compounds \cite{nomura2015}.
In DMFT calculations, the
relevant information on the kinetic term 
is the density of states (DOS). We 
choose $\ep_{\bm k\gm\gm'}=\ep_{\bm k}\delta_{\gm\gm'}$ and a featureless semi-circular DOS defined by 
$\sum_{\bm k}\delta(\ep-\ep_{\bm k}) = (4/\pi W^2)\sqrt{W^2-4\ep^2}$,
where $W$ ($\simeq$ 0.4 eV in fullerides \cite{nomura2015}) is used as the unit of energy. 
This DOS allows us to reveal the generic effect of negative Hund couplings and to discuss 
phenomena which are independent of material-specific details.

To study the symmetry-breaking transitions, we first truncate the interaction to the density-density components
of the form $n_{i\gm\sg}n_{i\gm'\sg'}$ with $n_{i\gm\sg}=c^\dg_{i\gm\sg}c_{i\gm\sg}$.
With this approximation, an anisotropy is introduced in spin and orbital space and the relevant quantities are $S^z$, $\tau^3$ and $\tau^8$.
Neglecting the spin-flip term 
$c^\dg_{i\gm \ua}c_{i\gm \da}c^\dg_{i\gm' \da}c_{i\gm' \ua}$ ($\gm \neq \gm'$)
is justified for $J<0$ since the electrons tend to occupy the same orbital and the 
inter-orbital spin degree of freedom is inactive.
On the other hand, dropping the pair hopping term 
$c^\dg_{i\gm \ua}c^\dg_{i\gm \da}c_{i\gm' \da}c_{i\gm' \ua}$ ($\gm \neq \gm'$)
is motivated by numerical simplifications,
 and we will discuss the effect of this approximation later.
The essential properties are correctly revealed by the simplified model with only density-density type interactions.

\begin{figure}[t]
\begin{center}
\includegraphics[width=85mm]{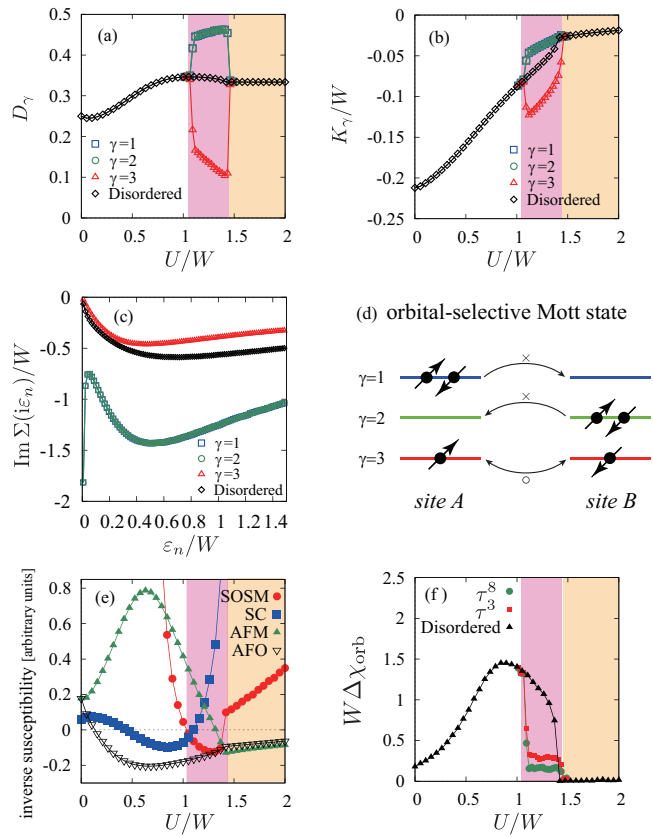}
\caption{(Color online)
(a) Orbital-dependent double occupancy and (b) kinetic energy as a function of $U$ at the temperature $T/W=0.005$ (results for a metallic initial solution).
Here we consider only density-density type interactions.
The highlighted regions show the SOSM state (pink) and Mott insulator (orange).
(c) Self energies at $U/W=1.25$
and $T/W=0.0025$.
(d) Schematic illustration of the SOSM state.
(e) Inverse susceptibility for AFM, SC, SOSM, and AFO orders calculated in the normal state without SSB, and (f) local orbital fluctuations at $T/W=0.005$.
The data in panel (e) have been rescaled, since we are interested only in the divergent points.
}
\label{fig:occ_kin}
\end{center}
\end{figure}

{\it Spontaneous orbital-selective Mott transition. ---}
To demonstrate the existence of the SOSM state, we consider a half-filled system, restrict the interactions to 
density-density type,
and set 
$J=-U/4$. This is a large Hund coupling compared to the realistic estimates for fulleride compounds \cite{nomura2012}, but this parameter choice allows us to clearly reveal the physics with modest computational resources. The results are qualitatively unchanged if we choose smaller Hund couplings, as shown in the SM \cite{supplementary}.
In the entire parameter regime considered in this study, the ordinary (uniform) orbital moment  
$\la \tau_i^\eta \ra$ 
is zero. 
To detect the transition to the SOSM state, let us introduce the orbital-dependent  
double occupancies and kinetic energies 
$D_\gm = \sum_i \la n_{i\gm\ua} n_{i\gm \da} \ra /N$
and 
$K_\gm = \sum_{\bm k\sg} \ep_{\bm k} \la c^\dg_{\bm k\gm\sg} c_{\bm k\gm\sg} \ra /N$,  
with 
$N=\sum_i 1$,
and plot these quantities as a function of $U$ (Fig.~\ref{fig:occ_kin}(a,b)). 
In the interaction range $1\lesssim U/W\lesssim 1.4$ 
 a spontaneous symmetry breaking is observed, which manifests itself 
in an orbital imbalance of $D_\gamma$ and $K_\gamma$.
If we focus on $D_\gamma $ and define the composite or two-body orbital moments ${\cal T}^{3,8} = \sum_\gm \lambda_{\gm\gm}^{3,8}D_\gm$, we have ${\cal T}^3=0$ and ${\cal T}^8>0$ in the present SOSM state (see also SM2 \cite{supplementary}).

To demonstrate the metallic and insulating characters of the different orbitals, we plot in panel (c) the purely imaginary self energies $\Sigma(\imu \ep_n)$ with $\ep_n =(2n+1)\pi T$ in the SOSM state.
Orbital $\gm=3$ behaves as a Fermi liquid ($\imag\Sigma(\varepsilon_n\rightarrow 0)\rightarrow 0$), while the self energies of the other orbitals diverge at low frequency, indicating a Mott insulating behavior.
From studies of the 
Hubbard model \cite{georges1996}, it is known that such a divergence of the self-energy is a characteristic feature of localized or paired electrons \cite{capone2002,micnas1990,koga2015}. 
A schematic illustration for the SOSM state is shown in panel (d).

The half-filled three-orbital system also exhibits conventional forms of 
symmetry breaking which can be detected by the divergence of the corresponding lattice susceptibilities, as shown in Fig.~\ref{fig:occ_kin}(e). 
In particular, there is an instability to 
intra-orbital-spin-singlet superconductivity (SC, order parameter $\la c^\dg_{i\gm\ua} c^\dg_{i\gm\da} \ra$) in the metallic region, and, if we assume a bipartite lattice to simulate bcc fullerides, antiferromagnetic order (AFM, order parameter $\la S^z_i \ra$) is observed in and near the Mott phase. 
The negative susceptibility for antiferro-orbital order (AFO, order parameter $\la \tau_i^{3,8} \ra$) 
indicates the presence of orbital order.
This order is however 
not stable against the pair hopping, as discussed
later. 
The SOSM state can also be detected by an odd-frequency susceptibility as shown in the figure, which 
is explained in the SM4 \cite{supplementary}.

\begin{figure}[t]
\begin{center}
\includegraphics[width=85mm]{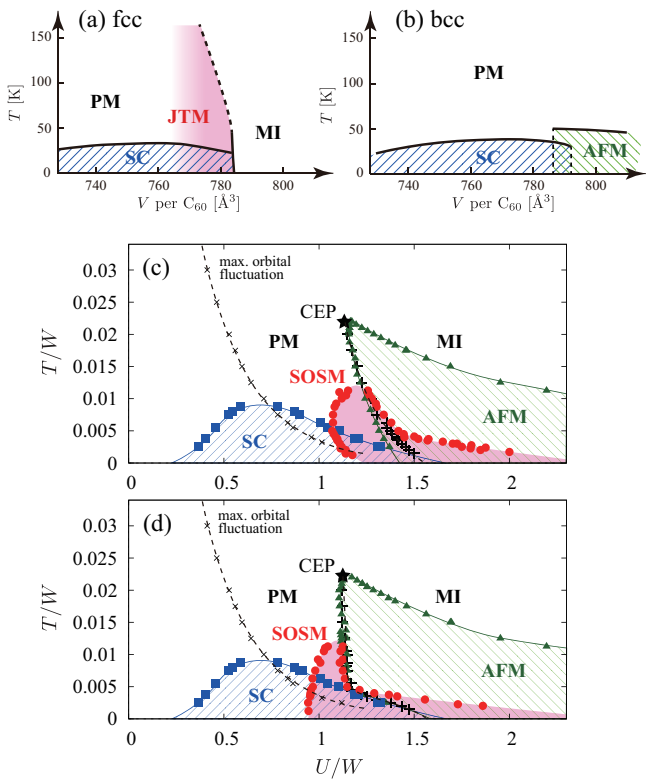}
\caption{
Sketches of the experimental phase diagrams for (a) fcc and (b) bcc fullerides based on Refs.~\onlinecite{takabayashi2009,zadik2015}.
Panels (c) and (d) show the phase diagrams of the three-orbital Hubbard model for density-density interactions and negative Hund coupling $J=-U/4$, calculated using a metallic and insulating initial solution, respectively.
The boundary of the SOSM phase has been determined from the points where the order parameter becomes finite, while the other boundaries are determined from the divergent points of the corresponding susceptibilities.
Here we do not consider the coexistence of two different orders. In overlapping regions the energetically favorable ordered state will dominate.
}
\label{fig:phase_all}
\end{center}
\end{figure}

By tracking these transition points we can map out the phase diagram as shown in Fig.~\ref{fig:phase_all}(c,d). For comparison, we also sketch the experimental phase diagrams for
fulleride compounds 
in panels (a) and (b).
The phase diagram obtained using a metallic initial solution is shown in panel (c) and the result for an insulating initial solution in panel (d). A hysteresis behavior is observed near the Mott transition point.
In both panels we indicate the Mott transition points of the system without SSB by black crosses. These critical points
delimit the stability regions of the 
paramagnetic metal (PM) and Mott insulator (MI), respectively, and end at a critical end point (CEP) at finite temperature. 
We note that it is difficult to accurately determine the metal-insulator transition points, 
since tiny numerical errors destroy the meta-stable insulating state near the boundary.

The SOSM state is stabilized near the Mott phase, while SC appears in the lower-$U$ region.
The transition into the SOSM and AFM phases is of first order: 
if the $U$-dependence is scanned from the metallic side 
the instabilities occur at different interaction values than if the calculation is started on 
the Mott insulator side.
The results also apply to the fcc lattice case if we neglect the AFM phase, which is sppressed by geometrical frustration.
Comparing the simulation results with the experimental phase diagrams sketched in Figs.~\ref{fig:phase_all}(a) and (b), 
we see that our model captures the characteristic properties of the fulleride compounds, including 
the dome shape of the SC region, and the Jahn-Teller metal, which is here identified as 
a 
SOSM state.
Although the Jahn-Teller metal has been observed only in the fcc system, our results suggest that it can be stabilized also on the bcc lattice.
At sufficiently low temperature, we observe the metal-insulator transition {\it inside} the SOSM phase.
However, 
this insulating SOSM state might not be detectable experimentally, since 
we expect 
magnetic order
at large $U$ and at low temperatures.

The SC dome is 
related to the local orbital fluctuations 
defined by 
$\Delta \chi_{\rm orb} = \int_0^{\beta} \diff \tau (\la \tau_i^\eta(\tau) \tau_i^\eta \ra - \la \tau_i^\eta (\beta/2) \tau_i^\eta \ra)$ with $\eta=3$ or $8$, 
where the second term is the long-(imaginary-)time correlator and $\beta=1/T$.
As shown in Fig.~\ref{fig:occ_kin}(f), 
these fluctuations reach a maximum at an interaction considerably lower than the Mott critical value. The temperature dependence of these maxima yields the crossover line marked with diagonal crosses in Figs.~\ref{fig:phase_all} (c,d). Evidently, the 
SC dome is peaked in the region of maximum orbital fluctuations. 
 In this sense, the intra-orbital spin-singlet pairing in fulleride compounds is the negative-$J$ analogue of the recently discussed fluctuating local spin-moment induced spin-triplet superconductivity in multiorbital systems with $J>0$
\cite{hoshino2015,steiner2016}.
More explicitly, the originally repulsive intra-orbital interaction $U$ can become attractive by considering the second-order perturbation contribution,
\begin{align}
\tilde U \simeq U - 4U'(U'+|J|)\chi_{\rm loc} + O(U^3),
\label{ueff}
\end{align}
as a result of enhanced 
local orbital fluctuations $\chi_{\rm loc}$ (see Refs.~\cite{inaba2002,hoshino2015}).
In the SOSM state, these local orbital fluctuations are suppressed, as shown in Fig.~\ref{fig:occ_kin}(f), and the results for the $\eta=8$ and $\eta=3$ components differ.
Although this indicates a suppression of SC, 
with the pair hopping term 
the fluctuations partly remain, and may even be enhanced 
as discussed below.

\begin{figure}[t]
\begin{center}
\includegraphics[width=80mm]{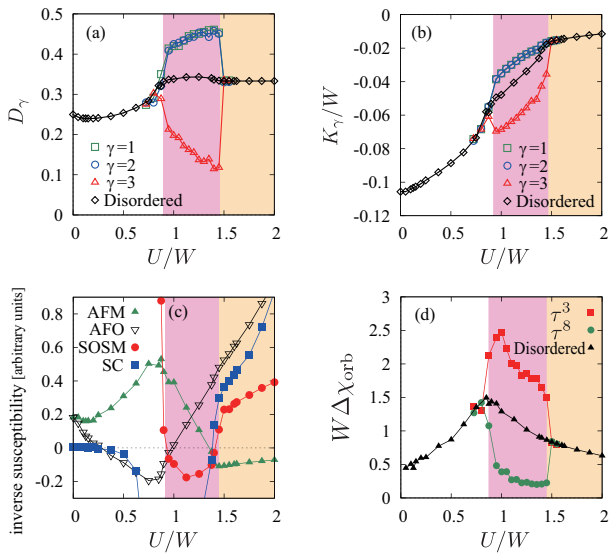}
\end{center}
\caption{
(Color online) 
Results analogous to Fig.~\ref{fig:occ_kin}, but for the model with full interaction including pair hopping. 
Here 
$J=-U/10$ and $T/W=0.005$.
}
\label{fig:pair}
\end{figure}

{\it Effect of pair hopping. ---}
To illustrate the effect of
the spin-flip and pair-hopping terms, which recover the continuous rotational symmetry in spin and orbital space, we present results for a Hund coupling of $J=-U/10$.
Figure \ref{fig:pair} shows the orbital-dependent double occupancies
and inverse susceptibilities for the AFM, AFO, SC and SOSM states. 
These quantities are the same as previously discussed for the density-density case in Fig.~\ref{fig:occ_kin}.
As shown in panel (a) of Fig.~\ref{fig:pair},
the double occupancies (and also kinetic energies) 
become orbital-dependent spontaneously for a range of interaction values near the Mott transition point.
Even though we choose here a smaller value 
of $|J|$
than in the density-density 
interaction, the widths of the SOSM regions in Figs.~1 and \ref{fig:pair} are almost identical.
This demonstrates that the pair hopping stabilizes the SOSM state.
A similar conclusion applies to SC, which covers a wider $U$-region than in the density-density case. 
We have also confirmed that the SOSM state can be stabilized for even smaller, and more realistic \cite{nomura2012} $|J|$ values, such as $J=-0.03U$ (see SM5 \cite{supplementary}).

The comparison of Fig.~\ref{fig:pair}(c) and Fig.~\ref{fig:occ_kin}(e) shows that 
the AFO order is 
suppressed due to the pair hopping term, which mixes different orbital states and washes out the orbital moments.
Hence the AFO order
in the model with only density-density type interactions
 is an artifact of the missing pair-hopping term.
Although there is a region where the AFO susceptibility is negative, this region is away from the Mott transition point.

As mentioned above, the spin-singlet pairing is strongly enhanced by the pair hopping, which means that it is underestimated in the 
model with only density-density interactions. 
This 
can be 
understood as follows: in the SC phase, the negative pair hopping term can be decoupled as
$|J| \sum_i c^\dg_{i1\ua}c^\dg_{i1\da} \la c_{i2\ua}c_{i2\da}\ra + {\rm H.c.}+ \cdots,$
which shows that condensation energy for pairing is gained 
at the static mean-field level, 
and 
it should increase the pair amplitude (and also $T_c$).
In the SOSM state,
orbital fluctuations of the $\tau^8$ moment are suppressed but those for $\tau^3$ are enhanced,
as shown in Fig.~\ref{fig:pair}(d). This is in contrast to Fig.~\ref{fig:occ_kin}(f).
Thus, in the rotationally invariant case, the orbital fluctuations persist and can help SC inside the SOSM state, compared to the case without pair hopping.
We note that it has been demonstrated that an imbalance of the Coulomb interaction, which produces an effect similar to the symmetry breaking field in the SOSM state, can enhance the superconductivity in the weak coupling regime \cite{kim2016}.
It is also notable that $\Delta\chi_{\rm orb}$ in the Mott insulator regime is enhanced compared to Fig.~\ref{fig:occ_kin}(f) due to the pair hopping.

We also comment on the origin of the SOSM state. The fluctuation of the composite moment ${\cal T}^8=(D_1+D_2-2D_3)/\sqrt 3$ at low temperatures 
can be captured already in the atomic limit with $U\rightarrow\infty$ 
(see the SM3 for more details \cite{supplementary}). Hence, the composite moment ${\cal T}^8$ tends to order in the lattice environment and should be regarded as the primary order parameter. On the other hand, the orbital imbalance of the kinetic energies, which appears 
simultaneously with ${\cal T}^8\ne 0$,
may be regarded as a secondary effect. This picture 
is further supported by the fact that the phase boundary of the (insulating) SOSM state has a long tail at large $U$ as shown in Figs.~\ref{fig:phase_all}(c,d), indicating the relevance of the $U\rightarrow \infty$ limit.

{\it Diagonal odd-frequency order. ---}
So far, we have characterized the 
SOSM state 
by the static orbital imbalance in the double occupancy and/or kinetic energy.
Here we show that both 
pictures are 
captured 
simultaneously 
by the 
time-dependent quantity 
\begin{align}
T^{\eta}(\tau) &= \sum_{i\gm\gm'\sg} 
\la c^\dg_{i\gm\sg} \lambda^{\eta}_{\gm\gm'} c_{i\gm'\sg}(\tau)\ra,
\label{eq:order_param}
\end{align}
which is odd with respect to time, since $T^{\eta}(0)$ is an ordinary orbital moment and thus vanishes.
If $T^{\eta}(\tau) $ becomes finite, it describes the orbital SSB.
To obtain a static representation of the order, we expand $T^{\eta}(\tau)$ with respect to imaginary time,
\begin{align}
T^{\eta}(\tau) &= 
T^\eta_{\rm even} + 
T^\eta_{\rm odd} \tau + O(\tau^2),
\label{Teta}
\end{align}
with zero ordinary order parameter: $T^{\eta}_{\rm even} = 0$.
The first-order 
contribution (or `odd-frequency' component) $T_{\rm odd}^\eta$, which characterizes the odd-time dependence of the order parameter of the SOSM state, may be regarded as a generalized orbital moment.
The explicit form is given by
\begin{align}
T_{\rm odd}^{8} = 
\sum_{\gm} \lambda^8_{\gm\gm} (K_\gm + 2U D_{\gm}) + (\rm other\ terms),
\label{eq:T18}
\end{align}
where 
the ``other terms'' originate from the interorbital interactions ($U'$ and $J$).
As seen in Eq.~\eqref{eq:T18}, $T_{\rm odd}^{8}$ 
incorporates
the orbital-dependent kinetic energies and double occupancies, and provides an alternative characterization of the SOSM state.
In a manner analogous to ordinary orders, the corresponding instability 
can be detected by looking for the divergence of the odd-frequency orbital susceptibility \cite{hoshino2011}, as shown in Figs.~\ref{fig:occ_kin}(e) and \ref{fig:pair}(c).
Hence this concept
provides 
a convenient tool for detecting the composite ordered phases.

{\it Summary and outlook. ---}
Orbital selective Mott states have been previously discussed in systems with an originally broken orbital symmetry, either due to orbital-dependent bandwidths \cite{anisimov2002,koga2004,inaba2006,medici2009} or crystal-field splittings \cite{werner2007}. The orbital selective states revealed in this work are of a fundamentally different nature, since they correspond to a {\it spontaneous} symmetry breaking in an orbitally degenerate system 
\footnote{A detection of collective excitation modes associated with the spontaneous orbital symmetry breaking could provide experimental evidence for the present SOSM phase.}.
This exotic state of matter is characterized by an unusual order parameter, i.e. an orbital-dependent double occupancy (composite order parameter) and/or kinetic energy. These physical quantities naturally appear in the time-dependent orbital moment with odd time dependence, which allows us to interpret the present order as an odd-frequency order. This concept, which has first been introduced as odd-frequency superconductivity, 
is generalized here to the diagonal-order version. Our work demonstrates the existence of this unconventional order in multiorbital Hubbard systems with negative Hund coupling, and provides strong evidence that it is actually realized in fulleride 
compounds, 
in the form of the Jahn-Teller metal phase. 

The SOSM states discussed here appear in half-filled multi-orbital systems with an odd number ($>1$) of orbitals. In the two-orbital case, the phenomenon cannot be observed, because the pairing of electrons will simply lead to a paired Mott insulator in all orbitals. If an odd number of electrons is present, as in the three orbital case, and also in half-filled five- or seven-orbital systems, there will always be at least one unpaired electron left, which can support metallicity.

There
exist several kinds of degenerate SOSM states, which are related by symmetry. 
It should thus be possible to switch from one SOSM phase to another by applying pressure, electric fields, or by photo-excitation. 
The experimental detection and control of these exotic ordered states will be an interesting future challenge. 
Since the different SOSM phases can be distinguished by their 
anisotropic conductivity, 
a reliable switching protocol could pave the way to energy efficient and fast memory devices.

{\it Acknowledgement. ---}
The authors thank A. Koga and Y. Nomura for valuable discussions.
S.H. acknowledges financial support from JSPS KAKENHI Grant Nos. 13J07701, 16H04021, and P.W. support from the ERC Starting Grant No. 278023.
The authors benefited from the Japan-Swiss Young Researcher Exchange Program 2014 coordinated by JSPS and SERI (Switzerland).
The numerical calculations have been performed using the supercomputer at ISSP (the University of Tokyo), the BEO04 cluster at the University of Fribourg, and the CET-7 cluster at CEMS (RIKEN).

\section*{SUPPLEMENTARY MATERIAL}

\subsection*{SM1. Orbital degrees of freedom and Gell-Mann matrices}
While spin-degrees of freedom can be described by the $2\times 2$ Pauli matrices, the orbital degrees of freedom are described by the Gell-Mann matrices.
We define the $3\times 3$ matrices by
\begin{align}
\lambda^0 &= 
\sqrt{\frac{2}{3}}
\begin{pmatrix}
1&& \\
&1& \\
&&1
\end{pmatrix}
,\ \ 
\lambda^1 =
\begin{pmatrix}
&1& \\
1&& \\
&&
\end{pmatrix}
,\ \ 
\lambda^2 = 
\begin{pmatrix}
&-\imu& \\
\imu&& \\
&&
\end{pmatrix}
,\nonumber \\
\lambda^3 &= 
\begin{pmatrix}
1&& \\
&-1& \\
&&
\end{pmatrix}
,\ \ 
\lambda^4 =
\begin{pmatrix}
&&1 \\
&& \\
1&&
\end{pmatrix}
,\ \ 
\lambda^5 = 
\begin{pmatrix}
&&\imu \\
&& \\
-\imu&&
\end{pmatrix}
,\nonumber \\
\lambda^6 &= 
\begin{pmatrix}
&& \\
&&1 \\
&1&
\end{pmatrix}
,\ \ 
\lambda^7 =
\begin{pmatrix}
&& \\
&&-\imu \\
&\imu&
\end{pmatrix}
,\ \ 
\lambda^8 = 
\frac{1}{\sqrt 3}
\begin{pmatrix}
1&& \\
&1& \\
&&-2
\end{pmatrix}.
\end{align}
Physically, these correspond to the charge $n=\lambda^0$, the angular orbital (or dipole) moment $(\ell_x, \ell_y, \ell_z) = (\lambda^7, \lambda^5, \lambda^2)$ and the quadrupole moment 
$(q_{x^2-y^2}, q_{3z^2-r^2}, q_{xy}, q_{yz}, q_{zx}) 
= (\lambda^3, \lambda^8, \lambda^1, \lambda^4, \lambda^6)
= (\ell_x^2-\ell_y^2, 3(\ell_z^2 - n^2), \overline{\ell_x\ell_y}, \overline{\ell_y\ell_z}, \overline{\ell_z\ell_x})$,  
where the overline symmetrizes the expression as $\overline{AB} = (AB+BA)/2!$.
The rank-1 tensors break time-reversal symmetry and couple to a magnetic 
field, while the rank-2 tensors are electric and couple to a strain field.
The orbital angular momentum 
$\bm L = \sum_{i\gm\gm'\sg}c^\dg_{i\gm\sg}\bm \ell_{\gm\gm'} c_{i\gm'\sg}$ 
behaves in a manner similar to the spin $\bm S$, and the total magnetic moment is given by 
the sum of these vectors.

\subsection*{SM2. Symmetry of the orbital moments}

\begin{figure}[t]
\begin{center}
\includegraphics[width=85mm]{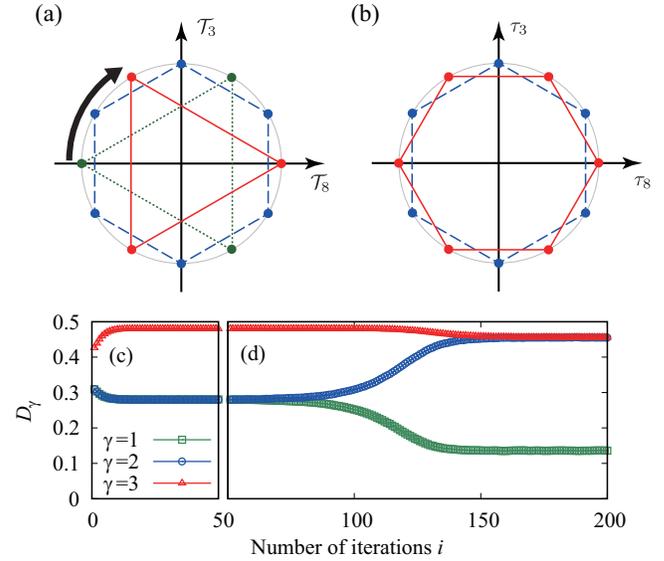}
\end{center}
\caption{
(Color online) 
Equivalent points in (a) the composite orbital moment ${\cal T}_8$-${\cal T}_3$ and (b) ordinary orbital momment $\tau_8$-$\tau_3$ planes at half filling, which are connected by lines with same color.
These points are related with each other by symmetry operations.
(c,d) Double occupancies as a function of number of iteration in the DMFT solution process. 
In (c) $\gm=1$ and $\gm=2$ are forced to be the same for the iteration number $i\leq 50$ and (d) there is no such constraint for $i>50$. 
Here we have chosen the parameters as $J=-U/4$, $U=1.25W$ and $T/W=0.005$ for the density-density type interaction.
The relaxation process in (d) is schematically shown as a thick arrow in (a).
}
\label{fig:orb}
\end{figure}

Here we examine the symmetries in the composite orbital moment (${\cal T}_8$, ${\cal T}_3$) space.
Since we have the double occupancies $D_\gm = \la n_{i\gm\ua} n_{i\gm \da} \ra$ in the three equivalent orbitals $\gm=1,2,3$, the Landau free energy, which is a scalar, must have the following form up to third order:
\begin{align}
F &= F_0 + a(D_1^2+D_2^2+D_3^2) 
\nonumber \\
&\ \ 
+ b(D_1D_2+D_2D_3+D_3D_1)
+cD_1D_2D_3
\nonumber \\
&\ \ 
+d[D_1^2(D_2+D_3)+D_2^2(D_3+D_1)+D_3^2(D_1+D_2)]
\nonumber \\
&\ \ 
+e(D_1^3+D_2^3+D_3^3)
+O(D_\gm^4)
,
\end{align}
where $F_0, a,b,c,d,e$ are constants.
Using the quantities 
\begin{align}
{\cal T}_0 &= \sum_{\gm} \lambda^0_{\gm\gm}D_\gm= \sqrt{\tfrac{2}{3}}(D_1+D_2+D_3),
\\
{\cal T}_3 &= \sum_{\gm} \lambda^3_{\gm\gm}D_\gm= D_1-D_2,
\\
{\cal T}_8 &= \sum_{\gm} \lambda^8_{\gm\gm}D_\gm= \sqrt{\tfrac{1}{3}}(D_1+D_2-2D_3),
\end{align}
this Landau free energy can be rewritten as
\begin{align}
F &= F_0' + a'({\cal T}_3^2+{\cal T}_8^2)
+ b' {\cal T}_8(3{\cal T}_3^2-{\cal T}_8^2)
+ O({\cal T}_\eta^4)
, \label{eq:free_energy2}
\end{align}
where the scalar part ${\cal T}_0$ is included in the coefficients.
The susceptibility in the disordered state is determined by the inverse second-order coefficients, which are identical for ${\cal T}_8$ and ${\cal T}_3$. 
If we define ${\cal T}_8 = r\cos \theta$ and ${\cal T}_3 = r\sin\theta$, the second-order term does not depend on the angle $\theta$, i.e. it is constant on the circle of radius $r$ in this plane. 
Hence the transition temperatures for these order parameters, which are determined by a divergence of the susceptibilities, are the same.
For any order parameter on the circle, the free energy of the ordered state becomes lower than the disordered state.

However, below the transition temperature the degeneracy between ${\cal T}_8$ and ${\cal T}_3$ is lifted by the third order term shown in Eq.~\eqref{eq:free_energy2}.
The third order term proportional to $b'$ may be rewritten in terms of $r$ and $\theta$, which yields the functional form $r^3 \cos 3\theta$.
Thus we only have a three-fold symmetry.
The present situation is illustrated in Fig. \ref{fig:orb}(a).

For comparison let us also consider the ordinary orbital moments.
In this case we replace $D_\gm$ by the one-body occupation $\la n_{i\gm \ua}+n_{i\gm \da}\ra$.
In contrast to the composite orbital moment discussed above, $\tau_\eta$ and its inverted component $-\tau_\eta$ are identical at half filling.
More specifically, by performing the particle-hole transformation we can change $\tau_\eta$ into $-\tau_\eta$ while keeping the Hamiltonian invariant.
Hence the Landau free energy must be an 
even function of $\tau_\eta$ and the third order term must vanish.
This situation is illustrated in Fig.~\ref{fig:orb}(b).
If we perform the particle-hole transformation for ${\cal T}_8$, on the other hand, it changes to ${\cal T}_8$ without any additional minus sign and the inversion symmetry is not present.

In order to seek the most stable solution in the ${\cal T}_8$-${\cal T}_3$ plane, we perform the following calculation. 
We first enforce the equivalence between the $\gm=1$ and $\gm=2$ components during the simulation, which stabilizes the ${\cal T}_8\neq 0$ state while keeping ${\cal T}_3=0$.
Figure \ref{fig:orb}(c) shows the iteration number $i$ dependence of the composite moments in the DMFT calculation, which converges for $i\leq 50$.
In this case, we have stabilized the ${\cal T}_8<0$ solution with ${\cal T}_3=0$.
After obtaining this solution, we release the above constraint of equivalent $\gm=1,2$ components, and continue the simulation further as shown in Fig.~\ref{fig:orb}(d) ($i>50$).
Then the solution slowly relaxes to another state, which is identical to the $ {\cal T}_8>0$ state. The corresponding evolution is schematically illustrated by the thick arrow in Fig.~\ref{fig:orb}(a).
Hence we conclude that the ${\cal T}_8>0$ state and its equivalent states are the most energetically stable.

\vspace{5mm}
\subsection*{SM3. Analysis of the atomic limit}

\begin{figure}[t]
\begin{center}
\includegraphics[width=60mm]{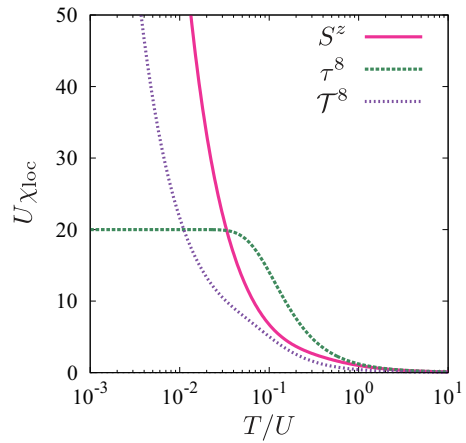}
\end{center}
\caption{(Color online) 
Temperature dependences of the local susceptibilities for spin ($S^z$), orbital ($\tau^8$), composite orbital moment $({\cal T}^8)$.
}
\label{fig:atom}
\end{figure}

To understand the origin of the SOSM state, let us consider the atomic limit by removing the kinetic energy term. Figure \ref{fig:atom}
shows the temperature dependence of the local susceptibilities of the spin ($S^z$) and conventional orbital moment ($\tau^8$). While the susceptibility for the spin moment $\bm S$ (and also the orbital angular momentum $\bm L$ which is not plotted here) show a divergent behavior at low temperatures, 
due 
to the six-fold ground state with $S=1/2$ and $L=1$ \cite{kim2016}, the one for the orbital moment $\tau^8$ 
(also for $\tau^{1,3,4,6}$)
does not.
However, the degrees of freedom associated with $\tau^8$ partially remain: the composite orbital moment ${\cal T}^8= (D_1+D_2-2D_3)/\sqrt 3$ 
shows a divergent behavior indicating the instability toward the (insulating) SOSM state.
Summing up these facts, we reach the conclusion that the composite order picture characterized by ${\cal T}^8$ is essential at least in the strong-coupling regime, and causes 
the orbital imbalance of the kinetic energy as a secondary effect in the presence of the inter-site electron hopping in a lattice environment.

The SOSM state is dominated by magnetic ($\bm S$ and $\bm L$) order 
at large $U$ and is not observed.
However it can dominate over 
those orders on the weak-$U$ side 
below the insulator-metal transition point. 
There, as demonstrated in Fig.~2 of
the main text, the SOSM state, 
which exhibits a mixture of metallic and insulating  behaviors,
is indeed realized.

\subsection*{SM4. Odd-frequency orbital susceptibility}

The composite orbital order is characterized by a two-body physical quantity as discussed in this paper.
Therefore the susceptibility which detects the instability is given by a four-body quantity, which is not easy to evaluate. 
It is thus more convenient to consider the instability of the first-order time derivative of the time-dependent orbital moment, which is known as the odd-frequency 
susceptibility \cite{jarrell1997,hoshino2011}.
We first introduce a homogeneous orbital moment operator by
\begin{align}
\mathscr{T}^\eta (\tau, \tau') = \sum_{i\gm\gm'\sg} c_{i\gm\sg}^\dg (\tau) \lambda^\eta_{\gm\gm'} c_{i\gm'\sg} (\tau') 
,
\end{align}
where we allow for $\tau\neq \tau'$.
For the ordinary orbital-ordered state, the order parameter is simply given by $\la \mathscr{T}^\eta (0,0)\ra$ $(=T_{\rm even}^\eta)$.
Let us assume that the orbital order occurs by a second-order phase transition, which is signaled by a divergence of the static susceptibility, 
\begin{align}
\chi^\eta_{\rm even} = \int_0^\beta 
\diff \tau \,
\chi^\eta(\tau,\tau,0,0)
, \label{eq:suscep_ordinary}
\end{align}
where 
the two-particle Green function 
is 
defined by
\begin{align}
\chi^{\eta} (\tau_1, \tau_2, \tau_3, \tau_4) &= \la {\rm T}_{\tau}
\mathscr{T}^\eta (\tau_1, \tau_2) \mathscr{T}^\eta (\tau_3, \tau_4)
\ra
,
\end{align}
with the time-ordering operator ${\rm T}_\tau$,
and its Fourier transform by
\begin{align}
\chi^{\eta} (\imu\ep_n, \imu\ep_{n'}) = \frac{1}{\beta^2} \int_0^\beta \diff\tau_1\cdots \diff\tau_4 \ 
\chi^{\eta} (\tau_1, \tau_2, \tau_3, \tau_4) 
\nonumber \\
\times \epn^{\imu\ep_n(\tau_2 - \tau_1)} \epn^{\imu\ep_{n'}(\tau_4 - \tau_3)}
.
\end{align}
Eq.~\eqref{eq:suscep_ordinary} is the ordinary susceptibility but is called ``even-frequency susceptibility'' to 
distinguish it from 
the ``odd-frequency susceptibility'' discussed below.
In terms of this quantity, Eq.~\eqref{eq:suscep_ordinary} can be expressed as
\begin{align}
\chi^{\eta}_{\rm even} = \frac{1}{\beta} \sum_{nn'} \chi^{\eta} (\imu\ep_n, \imu\ep_{n'})
.
\end{align}
This is a standard procedure to calculate susceptibilities.
In a similar manner, the other susceptibilities for magnetic 
and superconducting
orders 
can also be calculated.

We now consider the odd-frequency orbital moment defined in Eq.~(4) of the main text.
The instability toward the composite orbital order, or odd-frequency orbital order, can be detected by the correlation function
\begin{align}
&\int_0^\beta \la
\hat T_{\rm odd}^\eta (\tau) \hat T_{\rm odd}^\eta
\ra \, \diff \tau
= \chi^\eta_{\rm odd}
+ ({\rm regular\ part})
, \label{eq:oddfreq_suscep}
\\
&\chi^{\eta}_{\rm odd}
=- \frac{1}{\beta} \sum_{nn'} g(\ep_n) g(\ep_{n'}) \chi^{\eta} (\imu\ep_n, \imu\ep_{n'}),
\end{align}
where $g(\ep_n) = \ep_n \epn^{\imu \ep_n \eta}$ with $\eta = +0$, and $T_{\rm odd}^\eta = \la \hat T_{\rm odd}^\eta \ra$ which is defined in Eq.~(5) of the main text.
The regular term can be neglected as long as we are 
interested in 
the divergence of $\chi^{\eta}_{\rm odd}$.
The form factor $g(\ep_n)$ has appeared because of the time-derivative, and the factor $\epn^{\imu \ep_n \eta}$ is necessary for convergence. 
In the actual calculations, since we 
only consider the 
divergences of $\chi^{\eta}_{\rm odd}$, we can replace $g(\ep_n)$ by ${\rm sgn\,} \ep_n$, in which case we no longer need the convergence factor. 
The resulting susceptibility 
can correctly detect the instability toward odd-frequency orders with the order parameter $T_{\rm odd}^\eta$ \cite{hoshino2011} by its divergence.
The quantity $\chi^\eta_{\rm odd}$ is called ``odd-frequency susceptibility'' since it 
represents 
the odd-function part of the two-particle Green function with respect to frequency.

\vspace{10mm}
\subsection*{SM5. Additional data for realistic $|J|$}

\begin{figure}[b]
\begin{center}
\includegraphics[width=70mm]{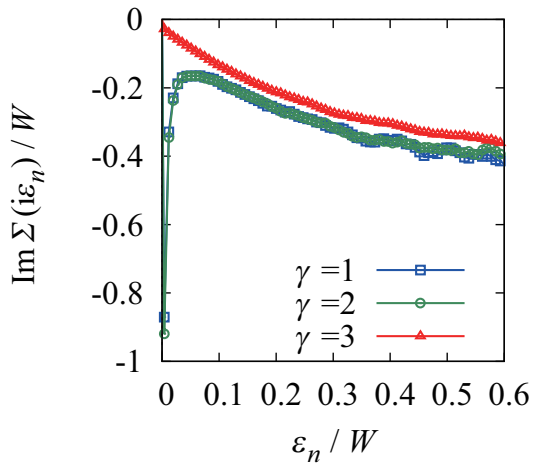}
\end{center}
\caption{(Color online) 
Self energy at $J/U=-0.03$, $U/W=1.325$, $T/W=0.00125$ for full interaction.
}
\label{fig:small_J}
\end{figure}

Figure \ref{fig:small_J} shows the self energy for a model with full interaction and $J/U=-0.03$.
This result demonstrates that the SOSM state is stabilized also for small negative Hund coupling, comparable in magnitude to the ab-initio estimates for the (Jahn-Teller screened) Hund coupling in A$_3$C$_{60}$ \cite{nomura2012}.

\end{document}